\documentclass{aa}
\usepackage{graphicx}
\usepackage{colortbl}
\usepackage{natbib}
\usepackage{amsmath}
\usepackage{txfonts}
\usepackage{adjustbox}

\begin{document} 
\title{X-ray linear polarization prediction in Black-Hole binaries and Active Galactic Nuclei and its measurements by IXPE}

\author{Lev Titarchuk\inst{1}
\and
Paolo Soffitta\inst{1}\thanks{corresponding author}
\and 
Elena Seifina\inst{2}
\and
Enrico Costa\inst{1}
\and 
Fabio Muleri\inst{1}
\and
Romana Mikusincova\inst{1}
}
\institute{INAF-IAPS, Via Fosso Del Cavaliere 100, 00133, Rome, Italy,\\
\email{lev.titarchuk@inaf.it, paolo.soffitta@inaf.it}
\and Lomonosov Moscow State University/Sternberg Astronomical Institute, Universitetsky Prospect 13, Moscow, 119992, Russia, 
\email{seif@sai.msu.ru}
}

   \date{received;accepted}

\abstract{We present a theoretical framework for the  formation of X-ray linear polarization in black hole (BH) source. The X-ray linear polarization originates from up-scatterings of initially soft photons within a hot, optically thick Compton cloud (CC) characterized by a flat geometry. We demonstrated that the degree of linear polarization is independent of photon energy and follows a characteristic angular distribution determined by the optical depth, $\tau_0.$ For $\tau_0$ $>5$, the linear polarization follows  the Chandrasekhar classical distribution. 
The  IXPE observations of several BHs X-ray binaries and of one Seyfert-1 galaxy confirm our theoretical prediction regarding values of the linear polarization $P$.} 
{The goal of this paper is to demonstrate that the main physical parameters of these Galactic and extragalactic sources can be derived without any free parameter using the polarization and X-ray spectral  measurements. These polarization measurements demonstrate that the polarization degree $P$ is almost independent of energy} 
{We estimate the CC optical depth, $\tau_0$ for all BHs observed by IXPE, using the plot of  $P$ vs $\mu$ = $\cos{i}$, where $i$ is an observer inclination  with respect to the normal, and considering the values $P$ for a given source.  Using X-ray spectral analysis, 
we obtain the photon index $\Gamma$ and, analytically determined  the CC plasma temperature $k_BT_e$.}
{Different BHs  in particular, Cyg~X--1, 4U~1630--47, LMC~X--1, 4U~1957+115,   
Swift~J1727.8--1613, GX~339--4, and the Seyfert-1 BH, NGC~4151 exhibited polarization in the 1\%--7$/$8\% level nearly independently of energy. $k_BT_e$ is in the range of 5--90 keV with a smaller value in High-Soft State with respect to Low-Hard State. Remarkably, a polarization vector parallel to the CC plane can be excluded solely based on spectral constraints, in agreement with the IXPE observations.}
{Using IXPE results on polarization and the inclination of the system we estimate $\tau_{0}$. Using  the photon index $\Gamma$ and $ \tau_{0}$,  we derive the plasma temperature $k_BT_e$ without any free parameters.  We find a  similarity between  the physical parameter 
and the IXPE findings and we provide evidence suggesting that the CC exhibits a flat geometry.}


\titlerunning{X-ray Polarization}
\maketitle


\section{Introduction}
Black hole (BH) X-ray binaries (BHXBs) evolve from the low/hard to the high/soft states (LHS and HSS states, respectively). 
In these states, the corona, presumably, dominates the X-ray emission and particularly in the LHS and the Intermediate States (IS) \citep{Montanari2009}. But the geometry of the corona is not  known yet.  The geometries  of the  corona considered up to now: 
a sphere above a BH as a lamp-post, 
\citep{Wilkins2012}, a Compton cloud (CC) as a flat (planer) atmosphere  \citep{Haardt1993} and a quasi-spherical CC between an accretiond disk (AD) and a central BH \citep{Shaposhnikov2006}. 

We return to the problem of the X-ray spectral formation in compact objects, in particular for a BHs, 
because of the quite exact measurements of the linear polarization for these objects; see e.g. \citep{Krawczynski2022, Steiner2024,Dovciak2024,Marin2024}.  
We remind  the reader details of the old paper by \cite{Sunyaev1985}, hereafter ST85, who investigated  analytically the X-ray spectral and linear polarization formation, $P$ in the slab (plain)  geometry)  for a wide range of the Thomson optical depth from 0.1 to more than 10. 
It is remarkable that, for an optical depth (half of the slab)  larger than 10, they obtained  a value of the polarization which follows the classical Chandrasekhar distribution versus the inclination $i$, that is the angle between the direction of observation and the normal to the slab \citep{Chandrasekhar1950}.
It is well known that the shape of the observed X-ray spectra agrees with the analytically derived Comptonization spectra \cite{Sunyaev1980, Titarchuk1994}, hereafter ST80 and T94, respectively. In this paper, it is important to remind a reader of the expected characteristics of the  linear X-ray polarization produced in a flat CC. 
We interpret the X-ray data from several Galactic Black-Hole binaries (BHXBs) and one active 
galactic nuclei (AGN) for which remarkable  polarization measurements have been obtained.

Stellar and supermassive black holes have similar accretion processes.
All these BHs show similar components in their spectra: a thermal component in the soft X-ray spectra, and a Comptonization component in the hard X-rays. 
The thermal component is likely generated within the accretion disk (AD), which is typically presented as a geometrically thin, optically thick structure, such as the Shakura-Sunyaev disk 
\citep[see][]{Shakura1973}. The Comptonized component is formed in the CC, a region of hot plasma that up-scatters photons from the accretion disk to higher energies via inverse Compton scattering, producing the hard X-ray emission observed in BHXBs and AGNs 
\cite[see][]{Titarchuk1994}.

In the HSS, the accretion disk blackbody emission dominates in the emergent spectrum,  see the X-ray spectral evolution in a BH XTE~J1650--500, \citep{Montanari2009}.  
Some authors can think that in the HSS, a geometrically thin optically thick accretion disk (AD), continues down to the innermost stable circular orbit \citep{Saade2024} while in the LHS the AD is possibly truncated. 
A cartoon showing the different geometry  for  a BH in the HSS and the LHS   is shown in Fig. 1
in ST85.

Polarization provides a direct constraint on the geometry of the X-ray-emitting region that is not possible by timing and spectral measurements, only. 
With this in mind, we applied this new independent way to derive relevant parameters of the accretion processes through the polarization data as demonstrated in ST85.
For the first time ever, the Imaging X-ray Polarimetry Explorer (IXPE) \citep{Weisskopf2022,Soffitta2021} a NASA-ASI mission with crucial INAF and INFN contributions , and launched on 9$^{th}$ December 2021, provides energy resolved X-ray polarization data  in the 2-8 keV energy band for a large number of stellar BHs \citep{Dovciak2024} in their different states and an handful Radio-Quiet AGNs \citep{Marin2024} .

In Section 2 we recall the polarization properties of the hard X-ray radiation produced  within  a planar CC around a BH.  
In Section 3, we review the polarization properties of the Galactic BHs and one bright AGN observed with the IXPE. 
In Section 4, we use the polarization properties of BHs to derive the main parameters of these objects.
In Section 5, we point out the importance of our results, and in Section 6, we draw the conclusions of our polarization study of the Galactic  and extragalactic BHs.

%
%
%
%

\begin{figure}
\raggedright
\includegraphics[scale=0.9,angle=0]{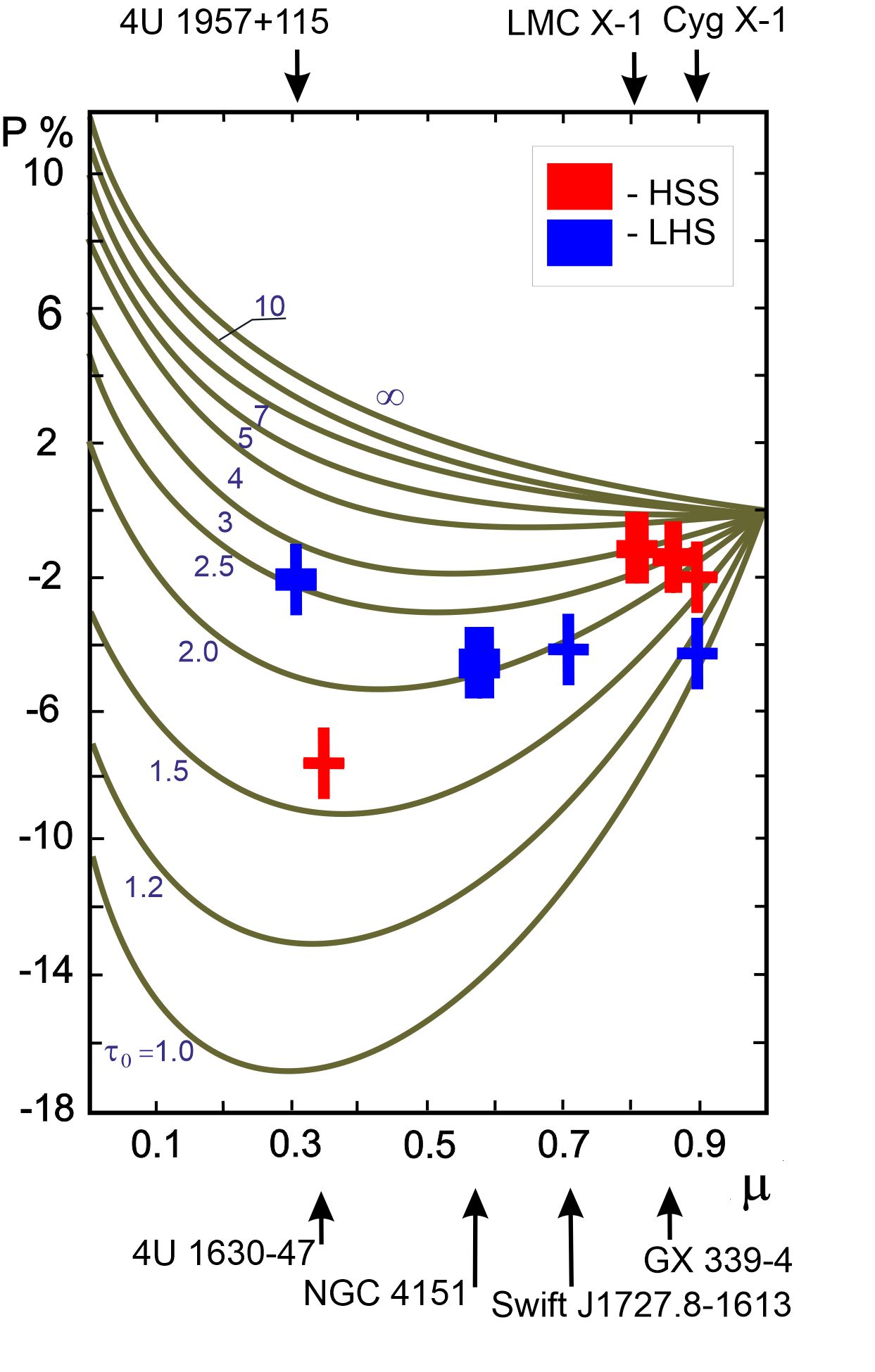}
\caption{The linear polarization P in \% as a function of $\mu$ =  $\cos i$, where $i$ is  the angle between the normal to the Compton cloud and a given direction (see ST85 and our Fig.~\ref{geometry_HSS_LHS}). 
Red  and blue boxes are drawn for the measured linear  polarization degree  
in the 2--8 keV for the soft and hard states correspondingly for different BH sources (source names are indicated by arrows next to the corresponding $\mu$). The vertical width of these boxes correspond to the error bars of the IXPE measurements. 
}
\label{pol_vs_mu}
\end{figure}

%

\begin{figure*}
\centering
\includegraphics[scale=1.45,angle=0]{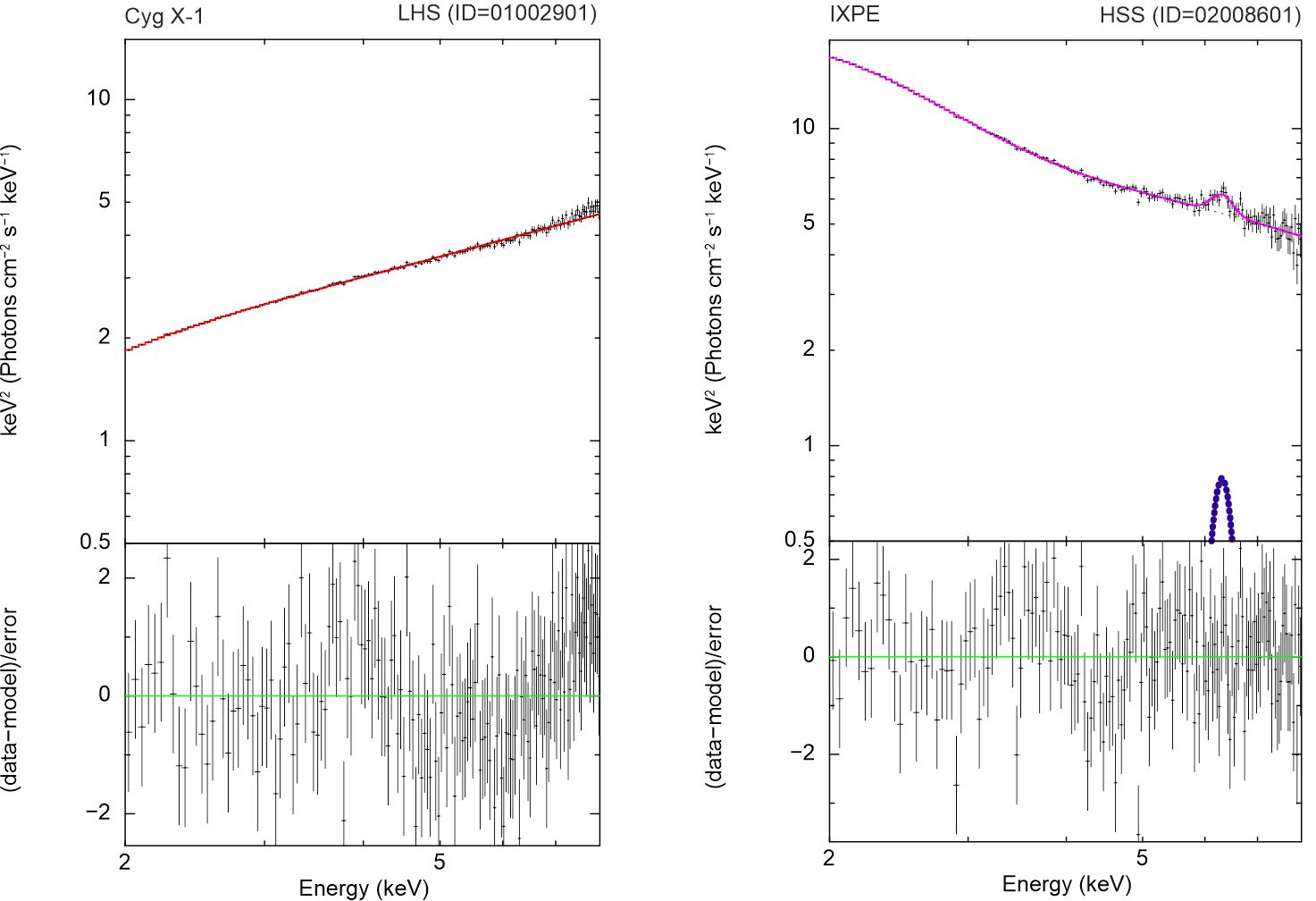}
\caption{The IXPE spectra of Cyg X-1 spectra for which X-ray  polarization is detected during the LHS and HSS. The photon index $\Gamma=1.4$ and $E_{line}=6.5$ keV, $\chi^2_{red}=158/149$, see left panel  (ID=01002901).
For  the HSS spectrum   these parameters are  $\Gamma=2.4$ and $E_{line}=6.4$ keV, $\chi^2_{red}=151/149$ see  right panel (ID=02008601). 
The best-fit model  consist of const*tbabs*(comptb + gaussian).}
\label{fig:CygX1LHHSSpectra}
\end{figure*}

%
%

\begin{figure}
\raggedright
\includegraphics[scale=0.5,angle=0]{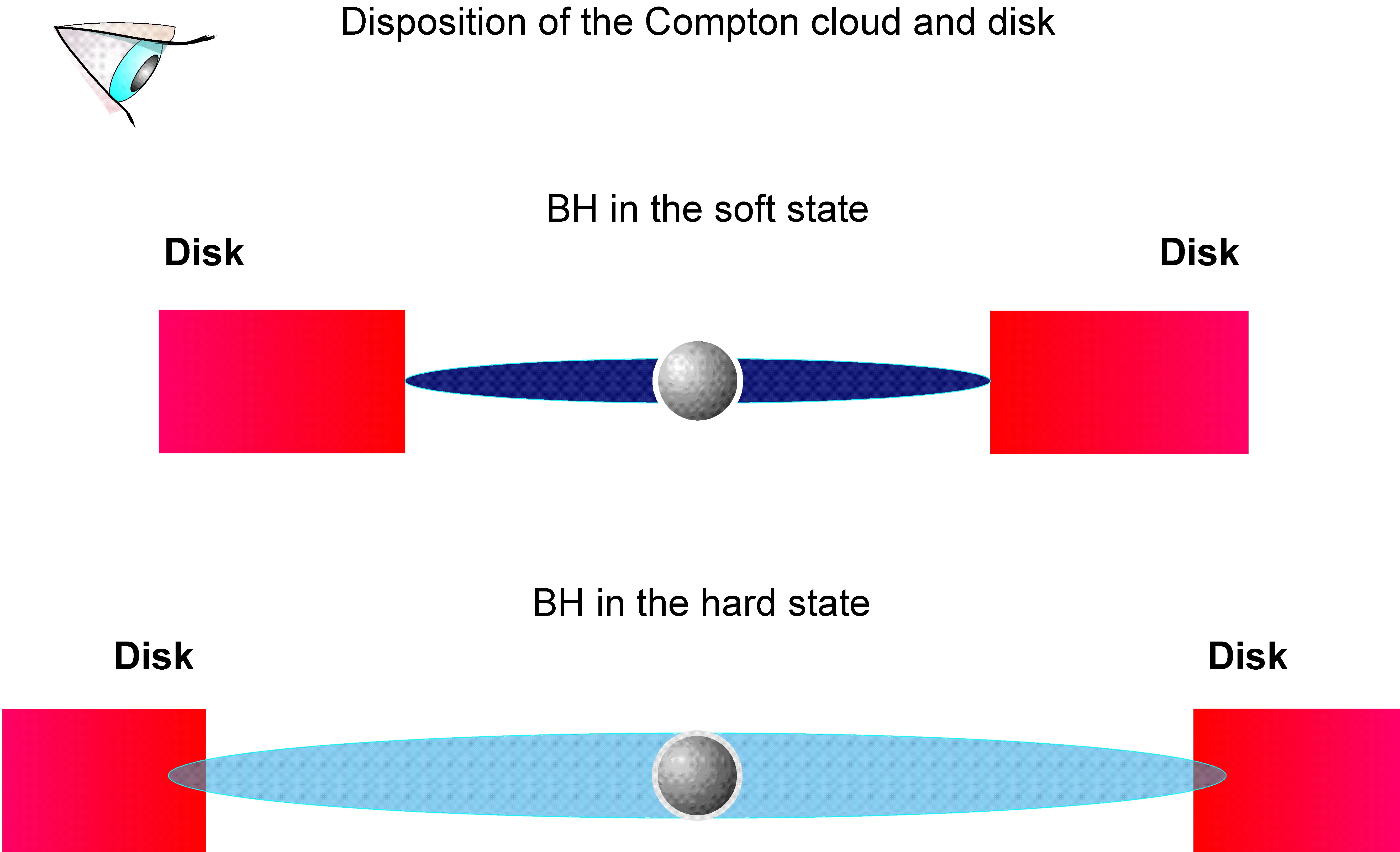}
\caption{Geometry of the Compton cloud.
Using the polarimetric measurement we assume  that
the planer Compton cloud (CC) is  present.
The CC is relatively smaller and thicker  for the soft state (dark blue) than that for the hard state (light blue).}

\label{geometry_HSS_LHS}
\end{figure}

%
%
 \section{The polarization X-ray photons emerging from Compton Clouds\label{polarization method}}
We assume that a primary source of the soft photons is distributed over the CC slab with different configurations:  in the center of the CC slab, uniformly distribution inside the Cloud or on the CC boundary.
ST85 showed that the polarization properties of photons that undergo a number of scatterings $N\gg\tau_0^2$, much larger than the average, for a given optical depth of the CC slab, $2\tau_0$ (see Fig.~1 in ST85), are independent of any of these distributions. 

Up-scattering (Comptonization) of low-frequency (soft) photons of $h\nu\ll k_BT_e$ leads to energy gain up to photon energies, $h\nu\le 3 k_BT_e$. 
A thermal, non-relativistic (i.e. Maxwellian) electron distribution is considered (see ST80, ST85).

It is a remarkable result of ST85 that the polarization of X-ray photons should be independent of the photon energy, which is essentially confirmed by the IXPE observations. Solving the integro$-$differential equation of the transport of the polarized  radiation, ST85 used the method of consecutive approximations. 
{\it For $k\gg\tau_0$ the polarization of photons after  these $k-$scatterings  is independent of $k$}, see ST85. When $k_BT_e< 70$ keV and the Thomson approximation is valid, the result of the ST85 calculations is correct \citep{Pozdnyakov1983}.  

We denote $I_l$ and $I_r$, we can refer them to as the two components vector $\bar{I} =(I_l, I_r)$, as the radiation intensities related to the electric field oscillations in the plane  defined by the disk axis  and by direction of photon propagation (plane called meridional) and in the plane  perpendicular, to it, respectively.
Changes of $I_l$ and $I_r$ in the CC plane have been calculated  by the following transport  equations  \citep{Chandrasekhar1950}:

\begin{equation}
\mu\frac{d}{d\tau} \begin{array}{c}I_l(\tau ,\mu)\\ I_r(\tau, \mu)\end{array}=\begin{array}{c}I_l(\tau, \mu)\\ I_r(\tau, \mu)\end{array} \int_0^1 P(\mu, \mu^\prime)
\begin{array}{c} I_l(\tau, \mu^\prime)\\ I_r(\tau,\mu^\prime)\end{array}  d\mu^\prime
-\begin{array}{c}F_l(\tau)\\ F_r(\tau),
\end{array}
\label{sc_matrix}
\end{equation}

where  $d\tau=-\sigma_TN_e dz$, $\mu=\cos i$ and 

\[ P(\mu, \mu^{\prime})=\frac{3}{8}
\left\{
\begin{array}{ll} 
2(1-\mu^2)(1-{\mu^\prime}^2)+\mu^2{\mu^\prime}^2
 & \mu^2 
 \\
{\mu^\prime}^2 & 1
\end{array}
\right \}  \]
 is the scattering matrix. The vector ${\bar{F}} =(F_l, F_r)$ notes  the distribution of primary sources. 
The boundary condition of the problem are:
\begin{equation}
{\bar{I}}(\tau=0, -\mu)= {\bar{I}}(\tau=2\tau_0, \mu)=0
\end{equation}
for any $0\leq \mu \leq1$.

The degree of the   polarization  is  defined as:
\begin{equation}
~~~~P=\frac{I_r-I_l}{I_r+I_l}
\label{pol_def}
\end{equation}
and the intensity is a sum $I(0,\mu)=I_r(0,\mu)+I_l(0,\mu)$.

We should find 
the photon distribution that undergoes $k-$ scatterings. One should also determine the distribution of photons which are not scattered in the medium,
namely:
\begin{equation}
{\bar{{ I}} (\tau, \mu)= \int_\tau^{2\tau_0} \exp[{ -(\tau^\prime-2\tau_0)/\mu)}]\bar{F}(\tau^\prime})d\tau^{\prime}/\mu 
\label{phot_no_scattered}
\end{equation}
for any initial photon distribution $\bar{F}(\tau)$. 

When the intensities of the first and next iterations  are calculated one can proceed with the intensity of photons which undergo $k-$scatterings:

\begin{equation}
\mu\frac{d}{d\tau} \begin{array}{c}I^k_l(\tau ,\mu)\\ I^k_r(\tau, \mu)\end{array}=\begin{array}{c}I^k_l(\tau, \mu)\\ I^k_r(\tau, \mu)\end{array}-\int_0^1 P(\mu, \mu^\prime)
\begin{array}{c} I^{k-1}_l(\tau, \mu^\prime)\\ I^{k-1}_r(\tau,\mu^\prime)\end{array}  d\mu^\prime.
\label{sc_matrix_by_itaration}
\end{equation}

The solution of the $k-$iteration depending of its solution on $(k-1)$ one,  see also ST85.

In Tab. 2 of ST85 the authors showed that the polarization degree and angle distribution 
of $k-$scattered photons  are independent of the primary source distribution if $k\gg\tau_0$.
To illustrate this statement the authors make their calculations for $\tau_0=2$ and  the iteration number $N\ge20$.
The calculation results for $k\gg\tau_0$  are independent of   $k$  and demonstrate that these results  can be applied  to the polarization of the Comptonized photons (see more details in ST85, Appendix A2).  It is worth to emphasize that for large optical depths  ($\tau\gg 5$) X-ray polarization  converges to the  classical solution by \cite{Sobolev1949} and \cite{Chandrasekhar1950}. The change of the polarization sign for $\tau< 4$ (see  Fig.~1 and ST85)  can be explained by a simple example of the optically thin CC (see \S 5.3 in ST85). 

The appropriate radiative transfer equation (1) can be rewritten in the operator form (see also \S A2 in ST85).  The system of equations of the polarized radiation (1) can be rewritten as 
 \begin{equation}
{\bar{I}}=L{\bar{I}}+{\bar{F}},
\label{operator_form}
\end{equation}
where $L$ is the  polarization integral operator.
It can be checked that the operator $L$ meets all the conditions   of the Hilbert-Schmidt (H-S) theorem, see ST85.
Thus, $L$ has a eigen$-$values ${\bar{p_i }}$ and 
ortho-normalized   eigen$-$functions ${\varphi}_i$ and $p_i\rightarrow0$ while $i\rightarrow\infty$.  A vector function ${\bar{F}}$  can be expanded in generalized Fourier series
 \begin{equation}
{\bar{F}}=\Sigma_{i=1}^{{\infty}} a_i {\bar{\varphi_i}}.
\label{vector_F}
\end{equation}

Solving Eq. (\ref{operator_form}) by the method of successive approximations we find the term ${\bar{I_k}}$ related to the photons  underwent $k-$scatterings:
 \begin{equation}
{\bar{I}}_k= \Sigma_{i=1}^{{\infty}} a_i p_i^k  {\bar{\varphi_i}}.
\label{I_k_term}
\end{equation}
Because of the sequence of $p_i$ decreasing with $k$  we obtain that 
 \begin{equation}
{\bar{I}}_k \simeq  a_1p_1^k{\bar{\varphi_1}}.
\label{final_form_I_k_term}
\end{equation}
This mathematical result is illustrated by Tab. 2 of ST85 for $k\gg \tau_0^2$. 

\section{Determination of the main parameters of the Compton Cloud  \label{spectral analysis}}
  
Sunyaev \& Titarchuk (ST80)  (see also ST85) solved the problem of X-ray spectral formation in the bounded medium. In particular, for the slab geometry they  derived a formula for the spectral index $\alpha$ (and $\Gamma=\alpha+1$):
  
 \begin{equation}
\alpha= -\frac{3}{2}
+\sqrt{\frac{9}{4}+\gamma},
 \label{sp_index}
 \end{equation}
 where $\gamma=m_ec^2\beta/k_BT_e$, $\beta=\pi^2/[12(\tau_0 +2/3)^2]$, $\alpha=\Gamma-1$ and $\tau_0$ is  the optical depth of a half of the slab. 
 
 Using these formulae one can find the electron temperature  $k_BT_e$ as a function of $\alpha$ and $\beta$:
 \begin{equation}
 k_BT_e=\frac{\beta m_ec^2}{(\alpha+3/2)^2-9/4}.
 \label{k_BTe_alpha_beta}
 \end{equation}

ST85 calculated the transfer of the polarized radiation using the iteration method. The number of iterations should be much more than the average number of scatterings in a given flat CC.  As a result they made a plot of the LP (linear polarization of the Comptonized photons) as a function of $\mu=\cos i$ where $i$ is an inclination  of the system for different Thomson optical depths of $\tau_0$. ST85 calculated the LP for $\tau_0$ from 0.1 to 10, and for $\tau_0>10$ they found that the LP follows the well-known Chandrasekhar distribution (see Fig.1).   
 
For quite a few BHs, the linear polarization $P$ was measured.
For example, for Cyg X-1 using IXPE data \citep{Steiner2024} showed that in the soft (HSS) and hard (LHS) states of the source, respectively,  $P_s\sim(2\pm0.5)\%$ 
$P_h\sim(3.5\pm0.5)\%$ are almost independent of the photon energy. The measured  photon indices  are: $\Gamma_{HSS}=2.5$,  $\Gamma_{LHS}=1.4$  and $k_BT_e=9$ keV, and  $90$ keV, respectively (see Table~\ref{tab:fit_table_asca_1957}).
 
 For the HSS of 4U 1630-47 \citep{Cavero2023} found that $\Gamma\sim 2.6-2.9$. The inclination $i$ is in the range of $60-75^\circ$, see \citep{Kuulkers1998}, and using the plot of the LP vs $\mu$, see Fig.~\ref{pol_vs_mu} we find that $\tau_0$ is around 1.7 and $k_BT_e\approx 10 $ keV [see Eqs.~(10), (11) and  Tab.~\ref{tab:fit_table_asca_1957}].

It is worth noting the trend of $P$, which ranges from 6$-$12\% 
in the 2$-$8 keV energy band as in  Tab. 1 \citep{Ratheesh2024}.
On the other hand, the photon indices $\Gamma$ in these measurements vary from 2.6 to 4.5, see \cite{Ratheesh2024}   and also \cite{Cavero2023}.
This indicates that the polarization detections were made while the source underwent significant spectral state changes.
It is important to emphasize that the quasi-constancy of $P$ with energy can only be ensured if the $P$ value is determined within the same spectral state. 

 For the BHXRB 4U~1957+115 \citep[][hereafter M24]{Marra2024} 
 obtained $P_h = (1.9\pm 0.6)\% $ which is observed in the HSS.  
 Knowing the inclination, $i\sim 72^\circ$ and using the LP vs $\mu$ plot, we find $\tau_0\sim 2.5$ for 4U~1957+115 ($\Gamma$ is about 2, see our Tab. \ref{tab:fit_table_asca_1957}). This is consistent with the result of M24, who found $\Gamma$ = 1.93$\pm$0.21 (Table 1 in M24).
 
For LMC~X--1 \citep[][hereafter P23]{Podgorny2023}  
reported an IXPE observation in the HSS  using NICER, NuSTAR and IXPE spectra (see their Fig.~3). The photon index, $\Gamma\approx2.6$  was obtained using the XSPEC NTHCOMP model.  P23 assumed that $i$ 
is about 36$^\circ$.  Thus, using  the P values and $\mu=\cos 36^\circ\sim$ 0.8 we find that the optical depth $\tau_0$  is about  2.5.
 
For Swift~J1727.8--1613 \citep{Veledina2023},  hereafter V23  reported a significant  polarization  [$P= (4.1\pm 0.2)\% $] which is almost independent of the photon energy (see Table~1 in V23). If we  take into account that $i$ is within the interval $30^\circ -60^\circ$  \citep{Cesares2014}, then we obtain  $\tau_0\sim 1.5$. V23 also reported that $\Gamma\sim1.8$ determined for a particular observation when the LP is about 4 $\%$.  In this case 
the optical depth, $\tau _0$ is about 2. 

For GX~339--4 \citep{Mastroserio2025} published a significant ($4\sigma$) polarization degree $P = (1.3 \pm 0.3)\%$ in the 3$–$8 keV band (where the spectrum is dominated by the corona) . We assume the inclination in this source to be $30^\circ$ as a low limit, see \citep{Mastroserio2025} Tab. 3. Thus, we get a value of $\tau_0\approx3$.

IXPE observed an handful of Radio-Quiet AGNs. One of the brightest Seyfert-1 galaxy, NGC~4151 showed a significant polarization $P=(4.5\pm0.9)\%$. parallel to the radio emission. If we obtain $\tau_0\approx2$ using Fig.~\ref{pol_vs_mu} using the inclination $i$ within the range of  
$58.1^\circ\pm0.9^\circ$ \citep{Bentz2022}.

%
%
\begin{table}
  \raggedright 
 \caption{BHXRBs and AGN  parameters including the best-fit photon index $\alpha$, the derived  optical thickness $\tau_0$ and the CC temperature  $k_BT_e$ }
 \begin{tabular}{lcccccclrcccccc}
      \hline
Source   & Spectral  &$\alpha=\Gamma-1$$^{\dagger}$   & $\tau_0$&  $k_BT_e$  (keV)   \\
         & state & & & \\
\hline 
Cyg X--1 & HSS & 1.4 & 2&  9 \\
Cyg X--1 & LHS & 0.4 & 1.2& 90   \\
\hline 
4U 1630--47 & HSS & 1.6 & 1.7&  10 \\
\hline 
4U 1957+115 & IS & 1.0 & 2.5&  10 \\
\hline
LMC X--1 & HSS & 1.6 & 2.5 & 6 \\ 
\hline 
Swift J1727.8--1613 & LHS & 0.8 &1.5 &  30 \\
\hline
GX 339--4& HSS & 1.4& 3& 7\\
\hline
NGC 4151& LHS & 0.85 & 2& 18 \\
\hline 
      \end{tabular}
      \label{tab:fit_table_asca_1957}
\tablefoot{ 
$^{\dagger}$ Here we used the Comptonization spectral model (ST80, ST85) 
(see comments in the text). 
}
\end{table}

\section{Discussion \label{results_discussion}}

One can argue on the sign of the observed  polarization degree (see  Fig.~\ref{pol_vs_mu}). 
If one chooses the  positive values
then he/she obtains that the inferred value, $\tau_0>5$ for almost any value of 
 $\mu=\cos i$. 
 But the derived value of 
$\tau_0> 5$  contradicts to the observed value of the index, 
$\Gamma> 1.5$ ($\alpha>0.5$) or the  Comptonization parameter $Y\sim( k_BT_e /m_ec^2) (2 \tau_0)^2\propto \alpha^{-1}$ 
(see Table \ref{tab:fit_table_asca_1957}). 
For these values of $\tau_0>5 $ the $Y$-parameter is higher than 2. 
In this case the emergent spectrum  has the Wien shape 
\citep[see e.g.,][]{Rybicki1979}, which shape  contradicts to the  X-ray observations.

We recall again here the definition of polarization as described by ST85. Negative polarization of the observed radiation implies that the polarization vector lies within the meridional plane, which is the plane defined by the disk normal and the propagation direction of the outgoing photons. Indeed, the IXPE observations of BHXRBs and Seyfert-1 galaxies show that the polarization vector is consistently parallel to the disk axis for all of them 
\citep[see, e.g.,][]{Krawczynski2022,Dovciak2024, Marin2024}. This confirms that the polarization vector resides within the meridional plane, explicitly ruling out polarization vectors oriented parallel to the disk plane.

The plane geometry is unexpected for the  CCs. As we noted in 
the Introduction that any type of  the CC geometry could be  assumed: jets above the disk or a  quasi-spherical   CC around a BH.  But the strong linear polarization can only  be if there is a strong asymmetry  in the CC around  a  BH.  It is natural  that for  the spherical  CC around the central object (e.g. a NS) the polarization degree 
is close to zero 
\citep[see e.g.,][]{Farinelli2024}.  If we choose the negative values of the polarization, $P$ then  it leads to values of $\tau_0$ from  0.1 to 3 (see Figs.~5 and 8 in ST85 and Fig.~\ref{pol_vs_mu} here). 

\citet[hereafter S24]{Saade2024} 
 makes a comparison of the X-ray polarimetric properties of 
 stellar and supermassive BHs. In their Fig.~1  they present  the polarization degree  vs their inclination $i$. The range of  $i$ for the Galactic BHs  is almost determined    while  extragalactic  ones  (EBH) have a very broad distribution (within approximately  the $60^\circ$ interval).  
 
 The PD values for these EBHs are typical  for those calculated in  ST85 and presented here in Fig.~\ref{pol_vs_mu}.   It is not by chance, that S24 noted that the polarization properties for these two  types  of sources are similar.  It is easy to see that the range of the optical depth $\tau_0$, in this case
 can be estimated  within  the interval of $1-1.5$.  If these  EBHs are observed  in the LHS then  $\Gamma<2$ and  the electron temperature is $k_BT_e \sim 40-60$ keV (see  Table \ref{tab:fit_table_asca_1957}).

 It is well known  that the best-fit values of the optical depth $\tau$ and $k_BT_e$ are around 2 and 60 keV,  respectively, in the LHS  and $\tau>3$, $k_BT_e< 10$ keV  in the HSS  (see Table 1 and e.g. ST85, T94).

\section{Conclusions}
\label{summary}
A  theoretical idea of the LP  formation in a BH  relies  on  the ST85 study.  The X-ray LP radiation is the result of the multiple  up-scattering   of the initially  X-ray soft photons (or UV ones  in the AGN case)   by the hot CC.  Multiple scattering  of the soft disk photons leads  to the  formation of the relatively hard emergent  Comptonization spectrum.  Furthermore, this specific spectrum is linearly  polarized if the CC  has  a flat geometry.  The polarization degree  of these multiple   up-scattered photons $P$  should be  independent of  the photon energy. 

ST85 made   a plot of  $P$  as a function of $\mu =\cos i$ where $i$ is  inclination of the source.   For  different  Thomson optical  depths  $\tau_0$  of the flat CC  ST85  calculated  the linear polarization $P$  
for $\tau_0$ from 0.1 to 10 (see also Figs.~5 and 8 in there).

IXPE observed quite a few  of  BHs and these observations  have been  already analyzed, see for example \citep{Dovciak2024}.
Cyg~X--1  data were studied by 
\citep{Krawczynski2022, Steiner2024} who  established that in the soft  and hard states  $P_s\sim2\%$ and  
$P_h\sim3.5\%$, respectively. 
\cite{Cavero2023} analyzed the data for 4U~1630--47 and they found that  
$P_h\sim 7\%$. \cite{Marra2024}  analyzed the data from a BH  source, 4U~1957+115.  
They  obtained  $P_h\sim 2\%$.

For  LMC~X--1 the analysis of the IXPE observations were made by \citep{Podgorny2023}. 
For this source IXPE found  that  $P$ is about $1\%$. 

For  4U~1630--47 \cite{Cavero2023} found  $P$  to be about $7\%$. 
  
 For Swift~J1727.8--1613 \citep{Veledina2023}  reported  
 that  $P\sim  4 \% $. 
 
For GX~339--4 the analysis of IXPE observations was made by Mastroserio et al. (2025). 
They found that $P\sim 1.3\%$.

For a bright Seyfert-1 galaxy (NGC~4151) 
\cite{Gianolli2024} found $P\sim  4.5 \%$.
 
For all these BHs using the plot of $P$ vs $\mu$  and values $P$ for a given source  we estimate the CC optical depth $\tau_0$ and using the results of  the spectral analysis,  we obtain  the photon index $\Gamma$,  the plasma temperature $k_BT_e$ without any free parameters (see Table 1).

IXPE observed a polarization vector aligned to the disk axis. This is consistent with the negative polarization inferred in this work based on the spectral shape analysis. A similar pattern is observed in radio-quiet Seyfert 1 galaxies studied by IXPE \citep{Marin2024}, highlighting a remarkable similarity across Black-Holes while spanning an enormous range in a BH mass \citep[e.g., see][]{Saade2024}.

\begin{acknowledgements}
The Imaging X-ray Polarimetry Explorer (IXPE) is a joint US and Italian mission.  The US contribution is supported by the National Aeronautics and Space Administration (NASA) and led and managed by its Marshall Space Flight Center (MSFC), with industry partner Ball Aerospace (now, BAE Systems).  The Italian contribution is supported by the Italian Space Agency (Agenzia Spaziale Italiana, ASI) through contract ASI-OHBI-2022-13-I.0, agreements ASI-INAF-2022-19-HH.0 and ASI-INFN-2017.13-H0, and its Space Science Data Center (SSDC) with agreements ASI-INAF-2022-14-HH.0 and ASI-INFN 2021-43-HH.0, and by the Istituto Nazionale di Astrofisica (INAF) and the Istituto Nazionale di Fisica Nucleare (INFN) in Italy.  This research used data products provided by the IXPE Team (MSFC, SSDC, INAF, and INFN) and distributed with additional software tools by the High-Energy Astrophysics Science Archive Research Center (HEASARC), at NASA Goddard Space Flight Center (GSFC). In addition, we acknowledge the fruitful discussion with the referee on the content of the presented paper.
\end{acknowledgements}

\bibliography{References} %
\bibliographystyle{aa} 

\end{document}